\documentclass[notitlepage]{article}

\usepackage{graphicx}
\newenvironment{acknowledgement}{Acknowledgements}

\begin{document}
\hyphenation{Czoch-rals-ki}

\author{S. Ganschow\footnote{Corresponding author, ganschow@ikz-berlin.de}, D. Klimm, P. Reiche, R. Uecker \\ Institut f\"{u}r Kristallz\"{u}chtung, Berlin, Germany}

\date{(Crystal Research and Technology \textbf{34} (1999) 615--619)}

\title{On the Crystallization of \\ Terbium Aluminium Garnet}

\maketitle

\noindent Attempts to grow terbium aluminium garnet (Tb$_3$Al$_5$O$_{12}$, TAG) by the \textsc{Czochralski} method lead to crystals of millimeter scale. Larger crystals could not be obtained. DTA measurements within the binary system showed that TAG melts incongruently at $1840^{\,\circ}$C. The perovskite (TbAlO$_3$, TAP) with a congruent melting point of $1930^{\,\circ}$C is the most stable phase in this system. The region for primary
crystallization of TAP covers the chemical composition of TAG and suppresses the primary crystallization of the terbium aluminium garnet.

\bigskip

Versuche zur Z\"{u}chtung von Terbium-Aluminium-Granat (Tb$_3$Al$_5$O$_{12}$, TAG) nach der \textsc{Czochralski} -- Methode erbrachten Kristalle
mit Gr\"{o}\ss en im Milli\-meter-Bereich. Die Z\"{u}chtung gr\"{o}\ss erer Kristalle mi\ss lang. DTA -- Messungen im bin\"{a}ren System ergaben,
da\ss\ TAG inkongruent bei $1840^{\,\circ}$C schmilzt. Der Perovskit (TbAlO$_3$, TAP) ist mit einem kongruenten Schmelz\-punkt von $1930^{\,\circ}$C die stabilste Phase in diesem System. Das Prim\"{a}raus\-scheidungsgebiet von TAP schlie\ss t die chemische Zusammensetzung des TAG mit ein und
unterdr\"{u}ckt die Erstkristallisation des Terbium-Alumi\-nium-Granats.

\section{Introduction}

\textsc{Faraday} isolators are a key component of many contemporary laser systems. They are used to prevent optical feedback leading to parasitic
oscillations in multi-amplifier systems and frequency instabilities in laser diodes. The most important optical element of a \textsc{Faraday} isolator is
the \textsc{Faraday} rotator providing a $45^\circ$ rotation of the polarization plane of traversing light. The characteristics that one looks for in a \textsc{Faraday} rotator material are a high \textsc{Verdet} constant, a low absorption coefficient, a low nonlinear refractive index and a high damage threshold. The most commonly used materials for the visible and near infrared regions are terbium doped glasses and terbium gallium garnet (TGG) single crystals. \textsc{Dentz} et al. (1974) reported the \textsc{Verdet} constant of TGG to be 134\,rad\,T$^{-1}$\,m$^{-1}$ at room temperature and for a wavelength of 633\,nm.

Already in the sixties \textsc{Rubinstein} et al. (1964) measured and published the \textsc{Verdet} constants of several rare earth aluminum garnets
that were grown from a lead oxyfluoride flux. Among these most promising for the application in \textsc{Faraday} devices is terbium aluminum garnet (TAG)
with a \textsc{Verdet} constant of 180\,rad\,T$^{-1}$\,m$^{-1}$ (room temperature, 633\,nm). This is caused by the electronic structure of the terbium ion i.e. the deep 5d bands and a wide range of transparency (\textsc{Weber} 1986). More than a decade later \textsc{Kuzanyan} et al. (1982) used a directional solidification technique to grow TAG. In their experiments they met with serious problems connected with the concurrent crystallization of TbAlO$_3$ (TAP, perovskite structure). These problems were explained by the action of impurities of the used terbium oxide, e.g. by an appreciable concentration of the higher oxide TbO$_2$ and a suspected lower melting point of the garnet phase.

Up to now, there are no reports in the literature about the successful growth of applicable TAG single crystals. Moreover, some of the information given
until now seems to be questionable e.g. concerning the melting temperature and the melting behavior.

\section{Experimental}

\subsection{Crystal growth experiments}

The stoichiometric mixture of the starting materials Al$_2$O$_3$ and Tb$_4$O$_7$ (5N and 4N5 purity, respectively) was sintered in a platinum crucible at 1450$^{\,\circ}$C for several hours in air. According to the relevant literature (e.g. \textsc{Levin, McMurdie} 1975) it can be assumed that TbO$_2$ contained in the reagents was entirely reduced to Tb$_2$O$_3$ before the material was molten. The reduction process Tb$^{4+}$ + e$^{-}$ $\rightarrow$ Tb$^{3+}$ becomes obvious by a change of color from brown to white.

Growth experiments were carried out applying an automated \textsc{Czochralski} technique with rf induction heating. \textsc{Kuzanyan} et al. (1982) reported for TAG a high melting temperature of 1860\ldots1900$^{\,\circ}$C. Therefore an iridium crucible was used and consequently, growth was carried
out under flowing nitrogen. In order to attain moderate temperature gradients and to avoid overheating of the melt an active afterheater was employed.
Crystallization from the crucible bottom was avoided by an active bottom heater. The growth parameters were rather common for oxides, i.e. a pulling
rate of 1.5\,mm/h was applied. A $\langle$111$\rangle$ oriented YAG crystal was used as seed.

In chapter \ref{chResults} it will be shown that all attempts failed to grow bulk crystals from the melt. It must be assumed, that this circumstance is due to peculiarities of the binary phase diagram Al$_2$O$_3$ -- Tb$_2$O$_3$.

\subsection{\label{ch_DTA}Differential thermal analysis}

The DTA measurements with 11 different Al$_2$O$_3$/Tb$_2$O$_3$ mixtures were carried out using a NETZSCH STA 409C equipment with tungsten high temperature TG-DTA sample holder (W/W-Re thermocouples). As the growth of TAG single crystals was the main intention of this study, only the relevant
part of the binary system Tb$_2$O$_3$ -- Al$_2$O$_3$ was investigated by DTA. The composition range under investigation (22\ldots59\,mol-\% Tb$_2$O$_3)$ included as well the garnet phase TAG (Tb$_3$Al$_5$O$_{12}$ --- $x_{\mathrm{Tb}_2\mathrm{O}_3}=0.375$) as the perovskite phase TAP (TbAlO$_3$ --- $x_{\mathrm{Tb}_2}\mathrm{O}_3=0.500$).

To get well homogenized DTA samples is was necessary to melt the material in a first heating cycle. The measurements were performed with sample masses of
typically 50\,mg under flowing argon and with heating/cooling rates of 10\,K/min. The onsets of the melting and crystallization peaks were observed to
coincide within a tolerance of typically 10\,K.

\section{\label{chResults}Results and discussion}

The crystal growth experiments did not result in TAG single crystals of satisfying size and quality. Typically only a few cubic millimeters of transparent and colorless material was grown directly on the seed and could be identified as TAG by X-ray powder diffraction analysis. Most of the
crystallized material was opaque of brownish color with lots of small spires on the surface. X-ray analysis confirmed that this part was a two-phase solid composed of TAG and TAP.

\begin{figure}[htb]
\begin{center}
\includegraphics[width=0.6\textwidth]{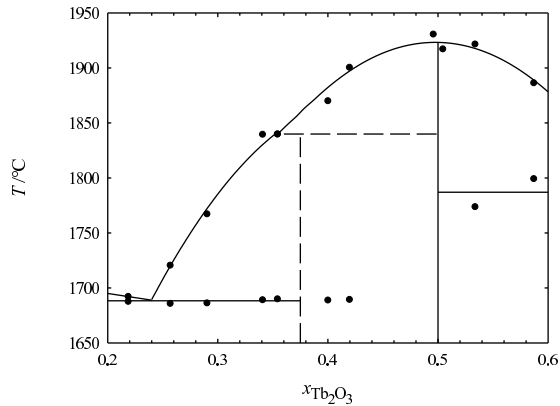}
\caption{DTA results within the Al$_2$O$_3$ -- Tb$_2$O$_3$ system and constructed binary phase diagram. The dashed and full vertical lines indicate
the compositions of TAG or TAP, respectively.}
\label{FigDiagram}
\end{center}
\end{figure}

According to these results one could assume that the congruent melting point of TAG simply does not coincide with the integer stoichiometric composition
but should be shifted toward lower terbium concentrations. Indeed, this assumption could explain the two phase crystallization mentioned above. After
the start with a stoichiometric TAG melt ($x=0.375$) lying between the two adjacent (congruently melting) phases the melt composition must follow the
liquidus line until the binary eutectic is reached.

\begin{figure}[htb]
\begin{center}
\includegraphics[width=0.7\textwidth]{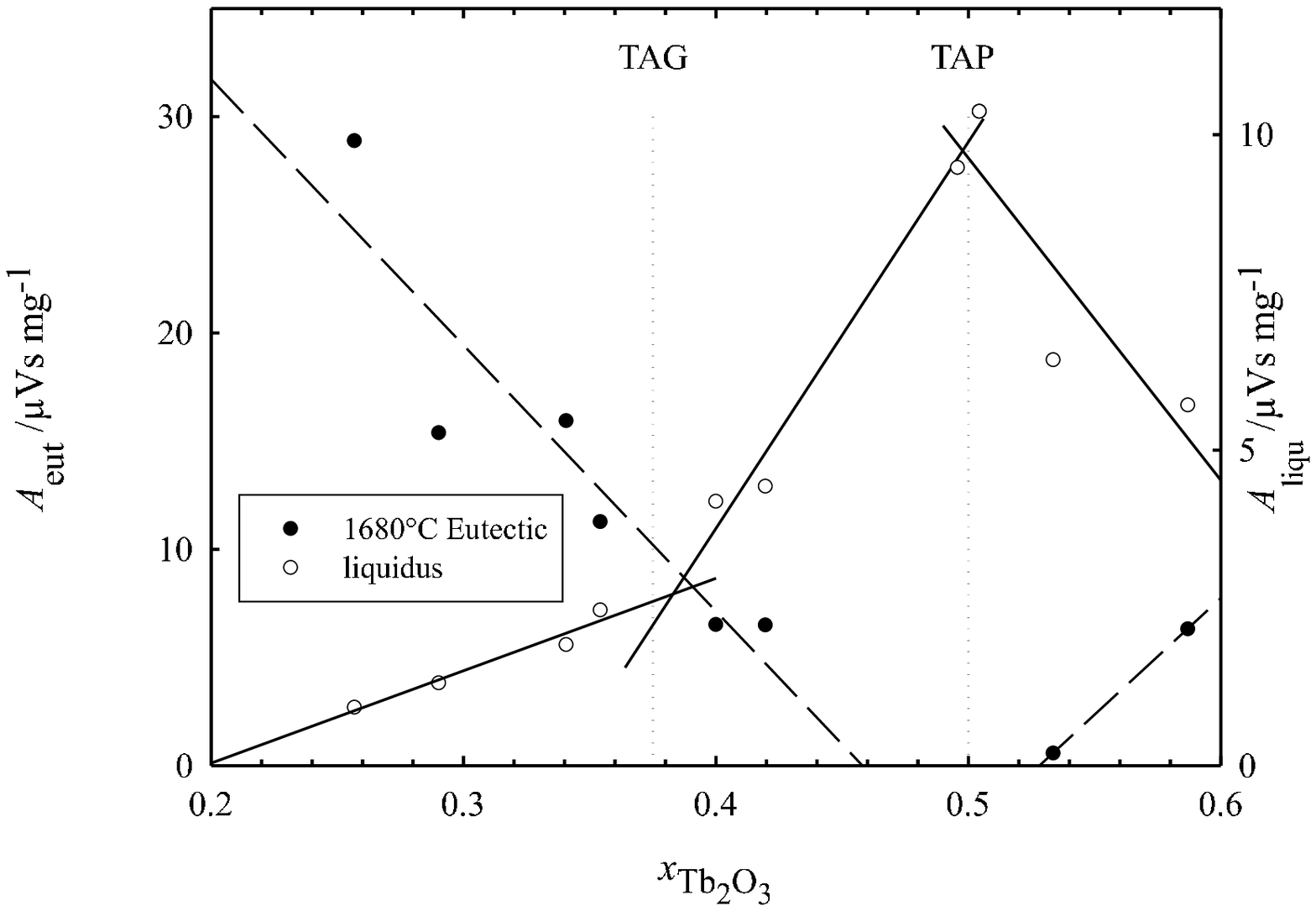}
\caption{Area of the liquidus peak size ($1692\ldots1931^{\,\circ}$C -- full line) and of the eutectic peak size ($1688^{\,\circ}$C left from TAP;
$\approx1790^{\,\circ}$C right from TAP -- dashed line) vs. $x_{\mathrm{Tb}_2\mathrm{O}_3}$.}
\label{FigPeakSize}
\end{center}
\end{figure}

The purpose of the DTA measurements within the system Al$_2$O$_3$ -- Tb$_2$O$_3$ that were reported in the previous chapter \ref{ch_DTA} was to find the expected congruently melting composition of TAG. The result was surprising since contrary to what had been reported (and believed) before. The expected liquidus maximum of TAG could not be found and hence, one has to take note of the fact that in the considered system the garnet phase is not a congruently melting one! Both heating and cooling curves could be used to construct the relevant part of the binary phase diagram (Fig.~\ref{FigDiagram}). The DTA results can be summarized as follows:

\begin{itemize}
\item For all 7 compositions with $0.22<x<0.42$ only one eutectic temperature $T_\mathrm{eut}^\mathrm{left}=(1688\pm2)^{\,\circ}$C was found. This is surprising, as on the left or right side of the composition TAG (Tb$_3$Al$_5$O$_{12}$, $x=0.375$) different eutectic temperatures for the eutectics Al$_2$O$_3$/TAG or TAG/TAP should be expected, if TAG was indeed a congruently melting compound. Only 2 compositions on the Tb$_2$O$_3$ rich side of TAP ($x=0.53$ and $x=0.59$) were investigated. Here an eutectic temperature $T_\mathrm{eut}^\mathrm{right}=(1787\pm15)^{\,\circ}$C could be found. The large experimental error is probably due to the small peak size, especially for $x=0.53$.

\item The liquidus shows only one maximum close to the composition TAP ($x=0.50$, $T_\mathrm{liqu}=1931^{\,\circ}$C. All liquidus temperatures for $0.4<x<0.59$ can be fitted by a parabola with apex at $x=0.498$ and $T_\mathrm{liqu}^\mathrm{max}=1923.3^{\,\circ}$C. The parabolic fit becomes much worse, if the liquidus temperatures for $0.257<x<0.354$ are included, as the slope $\partial T_\mathrm{liqu}/\partial x$ is flatter in this composition range left from TAG.

\item According to the lever rule it should be expected, that the peak areas of both eutectic and liquidus peak, $A_\mathrm{eut}$ and $A_\mathrm{liqu}$,
change linearly with $x$. An at least satisfying linear dependency could be found on both sides of TAP for $A_\mathrm{eut}$ (dashed lines in Fig.~\ref{FigPeakSize}). The observation, that from both sides $A_\mathrm{eut}=0$ is not reached at $x=0.50$ seems to indicate, that the compound TAP has a finite homogeneity range; but this hypothesis should be confirmed by additional investigations.
\end{itemize}

The separate linear fits of $A_\mathrm{liqu}$ at both sides of TAP are intersecting at $x=0.498$. This is exactly the same composition as obtained
for $T_\mathrm{liqu}^\mathrm{max}$, thus indicating, that this peak is indeed due to the melting/crystallization of TAP. From Fig.~\ref{FigPeakSize} it
becomes obvious, however, that the slope $\partial A_\mathrm{liqu}/\partial x$ changes from 61\,$\mu$\,V\,s/mg ($x>x_\mathrm{TAG}$) to 15\,$\mu$\,V\,s/mg ($x<x_\mathrm{TAG}$).

From the DTA measurements a peritectic decomposition temperature $T_\mathrm{per}=1840^{\,\circ}$C of TAG can be assumed. One could propose, that the observation of the eutectic peak at $T_\mathrm{eut}^\mathrm{left}=1688^{\,\circ}$C for 2 compositions with $x>0.375$ (cf. Fig.~\ref{FigDiagram}) indicates, that the chemical composition of the incongruently melting TAG is shifted to Tb$_{3+z}$Al$_{5-z}$O$_{12}$ ($z>0$). However, the observation of this peak for the sample with $x=0.42$ would result in the unexpected high value $z=0.36$. \textsc{DiGiuseppe} et al. (1980) reported for the gadolinium gallium garnet the congruently melting composition Gd$_{3.05}$Ga$_{4.95}$O$_{12}$. The corresponding value $z=0.05$ can be regarded typical for many garnets.

Accordingly, a better explanation for the $1688^{\,\circ}$C peak at $x>0.375$ seems the assumption, that the samples did not obtain the state of thermodynamic equilibrium during the DTA runs: If a melt with e.g. $x=0.4$ is cooled, crystallization of TAP takes place from $T_\mathrm{liqu}=1880^{\,\circ}$C down to $T_\mathrm{per}=1840^{\,\circ}$C. At this temperature TAP, TAG and a melt with $x\approx0.354$ are in
equilibrium and the peritectic reaction TAP + melt $\rightarrow$ TAG should consume melt. It must be expected, that a total equilibration can not be
obtained during the DTA runs due to the high cooling rate of 10\,K/min, as this reaction between a solid and a melt is diffusion controlled. Hence, parts of the melt remain and lead to the $1688^{\,\circ}$C peak due to eutectic crystallization.

\section{Conclusions}

The DTA measurements within the binary system Al$_2$O$_3$ -- Tb$_2$O$_3$ showed, that the incongruent melting behavior of TAG is responsible for the unsuccessful attempts to grow single crystals of this substance with considerable size. The small amounts of only a few cubic millimeter crystallized from an undercooled melt. Such crystallization seems possible, as at the composition of TAG ($x_{\mathrm{Tb}_2\mathrm{O}_3}=0.375$) the liquidus temperature is only $\approx15$\,K higher than $T_\mathrm{per}$.

The different thermodynamic stability and the resulting different melting/ de\-composition points of the garnet and perovskite phases in related
binary systems seem to be crucial problems for the growth of the garnets: YAG crystallizes easily from Al$_2$O$_3$/Y$_2$O$_3$ melts, as the melting
point of the garnet ($T_\mathrm{m}^\mathrm{YAG}=1970^{\,\circ}$C) is considerably higher than the melting point of the perovskite ($T_\mathrm{m}^{\mathrm{YAP}}=1875^{\,\circ}$C, \textsc{Abell} et al. 1974). The growth of GGG from Ga$_2$O$_3$/Gd$_2$O$_3$ melts is more difficult, as the garnet ($T_\mathrm{m}^\mathrm{GGG}=1740^{\,\circ}$C) melts below the perovskite ($T_\mathrm{m}^\mathrm{GGP}=1807^{\,\circ}$C). Here an overheating of the melts leads very often to the metastable crystallization of GGP even below $T_\mathrm{m}^\mathrm{GGG}$ (\textsc{DiGiuseppe} et al. 1980). The case of the Al$_2$O$_3$/Tb$_2$O$_3$ system presented in this work is even worse: The high thermal stability of the perovskite ($T_\mathrm{m}^\mathrm{TAP}=1930^{\,\circ}$C) is combined with peritectic decomposition of the garnet at $T_\mathrm{per}^\mathrm{TAG}=1840^{\,\circ}$C and prevents congruent crystallization of the garnet.

\bigskip

\begin{acknowledgement}
\newline The authors express their gratitude to M. Bernhagen for technical assistance.
\end{acknowledgement}

\bigskip

\renewcommand{\baselinestretch}{1.0}\textbf{References}

\noindent\textsc{Abell, J. S., Harris, I. R., Cockayne, B., Lent, B.}: J. Mat. Science \textbf{9} (1974) 527

\noindent\textsc{Dentz, D. J., Puttbach, R. C., Belt, R. F.}: Proc. AIP Conf. \textbf{18} (1974) 954

\noindent\textsc{DiGiuseppe, M. A., Soled, S. L., Wenner, W. M., Macur, J. E.}: J. Cryst. Growth \textbf{49} (1980) 746

\noindent\textsc{Kuzanyan, A. S., Ovanesyan, K. L., Petrosyan, A. G., Shirinyan, G. O.}: Dokl. Akad. Nauk Armyansk. SSR \textbf{74} (1982) 42

\noindent\textsc{Levin, M. McMurdie, H. F.}: Phase Diagrams for Ceramists, Suppl., Nat. Bureau of Standards, The American Ceramic Society (1975) (Fig. 4163)

\noindent\textsc{Rubinstein, C. B., Van Uitert, L. G., Grodkiewicz, W. H.}: J. Appl. Phys. \textbf{35} (1964) 3069

\noindent\textsc{Weber, M. J.}: SPIE Proc. \textbf{681} (1986) 75

\end{document}